\documentclass[prd,aps,twocolumn,preprintnumbers, showpacs, nofootinbib,superscriptaddress,notitlepage]{revtex4-1}
\usepackage{amssymb, amsthm}
\usepackage{amsmath}    
\usepackage{mathrsfs}   
\usepackage{graphicx}   
\usepackage{color}      
\usepackage{footnote}   

\usepackage{slashed}    
\usepackage{subfigure}  
\usepackage{hyperref}   

\usepackage{rotating}   
\usepackage{multirow}   
\usepackage[normalem]{ulem}

\newcommand{\non}{\nonumber}

\DeclareRobustCommand{\Eq}[1]{Eq.~\ref{#1}}

\begin{document}

\preprint{MSUHEP-18-004, MIT-CTP/5031}

\title{First direct lattice-QCD calculation of the $x$-dependence of the \\ pion parton distribution function}

\collaboration{\bf{Lattice Parton Physics Project (LP$\bold{^3}$)}}

\author{Jian-Hui Zhang}
\affiliation{Institut f\"ur Theoretische Physik, Universit\"at Regensburg, D-93040 Regensburg, Germany}

\author{Jiunn-Wei Chen}
\email{jwc@phys.ntu.edu.tw}
\affiliation{Department of Physics, Center for Theoretical Physics, and Leung Center for Cosmology and Particle Astrophysics, National Taiwan University, Taipei, Taiwan 106}
\affiliation{Center for Theoretical Physics, Massachusetts Institute of Technology, Cambridge, MA 02139, USA}

\author{Luchang Jin}
\affiliation{Physics Department, University of Connecticut,
Storrs, Connecticut 06269-3046, USA}
\affiliation{RIKEN BNL Research Center, Brookhaven National Laboratory,
Upton, NY 11973, USA}

\author{Huey-Wen Lin}
\email{hwlin@pa.msu.edu}
\affiliation{Department of Physics and Astronomy, Michigan State University, East Lansing, MI 48824}
\affiliation{Department of Computational Mathematics, Michigan State University, East Lansing, MI 48824}


\author{Andreas Sch\"afer}
\affiliation{Institut f\"ur Theoretische Physik, Universit\"at Regensburg, D-93040 Regensburg, Germany}


\author{Yong Zhao}
\affiliation{Center for Theoretical Physics, Massachusetts Institute of Technology, Cambridge, MA 02139, USA}


\begin{abstract}
We present the first direct lattice-QCD calculation of the Bjorken-$x$ dependence of the valence quark distribution of the pion. Using large-momentum effective theory (LaMET), we calculate the 
boosted pion state with long Wilson link operators.  
After implementing the one-loop 
matching and meson mass corrections, our result at $m_\pi \approx 310$~MeV is in agreement with those extracted from experimental data as well as from Dyson-Schwinger equation in small $x$ region, but a sizeable discrepancy in the large $x$ region. This discrepancy provides a nice opportunity to systematically study and disentangle the artifacts in the LaMET approach, which will eventually help to discern various existing analyses in the literature.
\end{abstract}
\maketitle

\section{Introduction}
The pion plays a fundamental role in QCD. As the lightest meson and the Goldstone boson associated with dynamical chiral symmetry breaking, it provides an important testing ground for our understanding of nonperturbative QCD. Currently, our experimental knowledge of the pion structure comes primarily from the Drell-Yan data for pion-nucleon/pion-nucleus scattering~\cite{Badier:1983mj,Betev:1985pf,Falciano:1986wk,Guanziroli:1987rp}. The valence quark distribution of the pion, $q_v^{\pi}(x)$ with $x$ being the fraction of the pion momentum carried by the active quark, has been extracted from these data~\cite{Conway:1989fs,Wijesooriya:2005ir,Aicher:2010cb}. Based on a next-to-leading order analysis including soft-gluon resummation, $q_v^{\pi}(x)$ was found to behave as $(1-x)^2$ at large $x$~\cite{Aicher:2010cb}. On the other hand, theoretical predictions of $q_v^{\pi}(x)$ have been made using various methods that are not fully consistent with this large-$x$ behavior. For example, the parton model~\cite{Farrar:1975yb}, perturbative QCD~\cite{Ji:2004hz,Brodsky:1994kg} and analysis from Dyson-Schwinger equations~\cite{Hecht:2000xa,Nguyen:2011jy,Maris:2003vk,Chen:2016sno} suggest that the valence quark distribution should behave like $(1-x)^a$ with $a\approx 2$, while relativistic constituent quark models~\cite{Frederico:1994dx,Szczepaniak:1993uq}, Nambu--Jona-Lasinio models~\cite{Shigetani:1993dx,Davidson:1994uv,Weigel:1999pc,Bentz:1999gx} and other arguments~\cite{Drell:1969km,West:1970av,Melnitchouk:2002gh} favor a linear dependence on $1-x$ at large $x$. For a review of the experimental and theoretical status of the pion parton distribution function (PDF), see Ref.~\cite{Holt:2010vj}. Lattice QCD should be able to shed light on this puzzling disagreement, provided that its computational potential can be extended beyond the first few moments of PDFs.

This became possible recently due to the breakthrough made by large-momentum effective theory (LaMET) in direct lattice calculation of the $x$-dependence of PDFs~\cite{Ji:2013dva,Ji:2014gla}. According to LaMET, the full PDF, instead of its first few moments, can be directly accessed from lattice QCD using the following method: 1) Construct an appropriate static-operator matrix element (now known as the quasi-PDF) that approaches the PDF in the large-momentum limit of the external hadron. The quasi-PDF constructed this way is usually hadron-momentum-dependent but time-independent, and, therefore, can be readily computed on the lattice. 2) Calculate the quasi-PDF on the lattice. 3) Convert it to the PDF through a factorization formula accurate up to power corrections suppressed by the hadron momentum. The existence of such a factorization is ensured by construction; for a proof, see Refs.~\cite{Ma:2014jla,Ma:2017pxb,Izubuchi:2018srq}.

LaMET has been applied to compute various nucleon PDFs~\cite{Lin:2014zya,Chen:2016utp,Lin:2017ani,Alexandrou:2015rja,Alexandrou:2016jqi,Alexandrou:2017huk,Chen:2017mzz} as well as meson distribution amplitudes (DAs)~\cite{Zhang:2017bzy,Chen:2017gck}, and yields encouraging results. The hard matching kernel appearing in the factorization of the quasi-PDF as well as quasi-DA has been computed in different schemes at one-loop order~\cite{Xiong:2013bka,Ji:2015jwa,Ji:2015qla,Xiong:2015nua,Ji:2014hxa,Monahan:2017hpu,Ji:2018hvs,Stewart:2017tvs,Constantinou:2017sej,Green:2017xeu,
Izubuchi:2018srq,Xiong:2017jtn,Wang:2017qyg,Wang:2017eel,Xu:2018mpf}, while the corresponding mass corrections are available in Refs.~\cite{Chen:2016utp,Zhang:2017bzy}. The renormalization property of the quasi-PDF has been investigated in Refs.~\cite{Ji:2015jwa,Ishikawa:2016znu,Chen:2016fxx,Constantinou:2017sej,Ji:2017oey,Ishikawa:2017faj,
Green:2017xeu,Chen:2017mzz}, and multiplicative renormalizability in coordinate space was established to all orders. In Refs.~\cite{Alexandrou:2017huk,Chen:2017mzz}, a 
nonperturbative renormalization in the regularization-independent momentum-subtraction (RI/MOM) scheme has been implemented. Certain technical issues regarding the nonperturbative renormalization as well as other aspects of the quasi-PDF, were also raised and addressed in Refs.~\cite{Constantinou:2017sej,Alexandrou:2017huk,Green:2017xeu,Chen:2017mzz,Chen:2017mie,Lin:2017ani,Chen:2017lnm,
Li:2016amo,Monahan:2016bvm,Radyushkin:2016hsy,Rossi:2017muf,Carlson:2017gpk,Ji:2017rah}. 
In the state-of-the-art calculation of the unpolarized isovector quark PDF~\cite{Chen:2018xof} (see also Ref.~\cite{Alexandrou:2018pbm}), the operator mixing at $O(a^0)$ with $a$ being the lattice spacing has been avoided, and the result agrees with the global PDF fit~\cite{Dulat:2015mca,Ball:2017nwa,Accardi:2016qay} within errors. Besides direct lattice calculations, there have also been studies of the quasi-PDF's and quasi-DA's using various models~\cite{Gamberg:2014zwa,Nam:2017gzm,Broniowski:2017wbr,Jia:2017uul,Hobbs:2017xtq} or using non-relativistic QCD in the heavy-quarkonium system~\cite{Jia:2015pxx}.
In parallel with the progress using the LaMET approach, other proposals to calculate the PDFs in lattice QCD have been formulated~\cite{Ma:2014jla,Ma:2017pxb,Radyushkin:2017cyf,Liu:1993cv,Liang:2017mye,Detmold:2005gg,Braun:2007wv,Bali:2017gfr,
Chambers:2017dov}, each of which is subject to its own systematics. However, these approaches can be complementary to each other and to the LaMET approach.

In this paper, we carry out the first direct lattice calculation for the valence quark distribution of the pion using the LaMET approach.
The calculation is done using clover valence fermions on an ensemble of gauge configurations with $N_f=2+1+1$ (degenerate up/down, strange and charm) flavors of highly improved staggered quarks (HISQ)~\cite{Follana:2006rc}  generated by the MILC Collaboration~\cite{Bazavov:2012xda}
 with lattice spacing $a = 0.12$~fm, box size $L \approx 3$~fm and pion mass $m_\pi \approx 310$~MeV. Our results are comparable quantitatively with the results extracted from experimental data~\cite{Aicher:2010cb} as well as from the Dyson-Schwinger equation~\cite{Chen:2016sno}.

\section{From quasi-PDF to PDF in the pion} 
The quark PDF in the pion is defined as
\begin{multline}\label{qpdf}
q_f^\pi(x) = \int\frac{d\lambda}{4\pi} e^{-i x \lambda n\cdot P} \\
 \times \left\langle \pi(P)\left|\bar\psi_f(\lambda n)\slashed n\Gamma(\lambda n,0)\psi_f(0)\right|\pi(P)\right\rangle,
\end{multline}
where the pion has momentum $P^\mu=(P_0,0,0,P_z)$, $\psi_f, \bar\psi_f$ are the quark fields of flavor $f$, $n^\mu=(1,0,0,-1)/\sqrt 2$ is a lightlike vector, $x$ denotes the fraction of pion momentum carried by the quark, and
\begin{equation}
\Gamma \left( \zeta n , \eta n \right) \equiv
  \exp \left( ig \int_\eta^\zeta d\rho\,n \cdot A(\rho n) \right)
\end{equation}
is the gauge link. The valence quark distribution is given by $q_{f,v}^\pi(x)=q_f^{\pi}(x)-q_{\bar f}^{\pi}(x)$ with $q_{\bar f}^{\pi}(x)=-q_f^{\pi}(-x)$, and satisfies $\int_0^1 dx\, q_{f, v}^{\pi}(x)=1$. For a charged pion, we have $q_{u,v}^\pi(x)=q_u^{\pi}(x)-q_{\bar u}^{\pi}(x)=q_u^{\pi}(x)-q_d^{\pi}(x)$ due to isospin symmetry.

The quark quasi-PDF can be defined in a similar way to \Eq{qpdf}:
\begin{multline}\label{quasipdf}
\tilde q_f^{\pi}(x) = \int\frac{d\lambda}{4\pi}e^{-i x \lambda \tilde n\cdot P} \\
  \times \left\langle \pi(P)\left|\bar\psi_f(\lambda \tilde n)\slashed {\tilde n}\Gamma(\lambda \tilde n,0)\psi_f(0)\right|\pi(P)\right\rangle,
\end{multline}
except that $\tilde n^\mu=(0,0,0,-1)$ is a spacelike vector with $\tilde n\cdot P=P_z$. As pointed out in Refs.~\cite{Xiong:2013bka,Hatta:2013gta}, the Dirac matrix $\slashed{\tilde n}=\gamma^z$ can also be replaced by $\gamma^t$, which has the advantage of avoiding mixing with scalar PDF~\cite{Constantinou:2017sej,Chen:2017mie}. The choice of $\gamma^t$ will be used throughout this paper.

The factorization connecting the quasi-PDF and the PDF was first presented in Refs.~\cite{Ji:2013dva,Xiong:2013bka} for bare quantities. Later on, it was shown~\cite{Chen:2016fxx,Ji:2017oey,Ishikawa:2017faj,
Green:2017xeu} that the renormalization factor of the quasi-PDF depends on the exponential of the Wilson line length times the Wilson line self energy counterterm.
Nonperturbative renormalization can be carried out by extracting the counterterm from the heavy quark potential~\cite{Chen:2016fxx,Zhang:2017bzy}, using the RI/MOM scheme~\cite{Alexandrou:2017huk,Stewart:2017tvs,Izubuchi:2018srq}, or by forming a ratio of the lattice matrix elements of the quasi-PDF at two different momenta~\cite{Radyushkin:2017cyf,Orginos:2017kos,Radyushkin:2018cvn,Zhang:2018ggy,Izubuchi:2018srq}. Following our previous studies of nucleon PDFs~\cite{Chen:2018xof}, we perform nonperturbative renormalization in the RI/MOM scheme, but we also show the result from Wilson line renormalization in intermediate steps for comparison and to estimate the size of systematic uncertainties.

In the RI/MOM scheme, the bare coordinate-space matrix element $\tilde h(\lambda\tilde n)$ showing up on the right hand side of \Eq{quasipdf} can be renormalized nonperturbatively by demanding that the counterterm $Z$ cancels all the loop contributions for the matrix element in an off-shell external quark state at a specific momentum~\cite{Stewart:2017tvs,Chen:2017mzz}: 
\begin{equation}\label{hR}
\tilde h_R(\lambda\tilde n)=Z^{-1}(\lambda\tilde n, p_z^R, 1/a, \mu_R) \tilde h(\lambda\tilde n), 
\end{equation}
and
\begin{align}\label{hRx}
&Z(\lambda\tilde n,p^R_z, 1/a, \mu_R)\nonumber\\
&=\left.\frac{\textrm{Tr}[\slashed p \sum_s \langle p,s|\bar\psi_f(\lambda \tilde n)\slashed {\tilde n}\Gamma(\lambda\tilde n,0)\psi_f(0)|p,s\rangle]}{\textrm{Tr}[\slashed p  \sum_s \langle p,s|\bar\psi_f(\lambda \tilde n)\slashed {\tilde n}\Gamma(\lambda\tilde n,0)\psi_f(0)|p,s\rangle_{tree}]}\right|_{\tiny\begin{matrix}p^2=-\mu_R^2 \\ \!\!\!\!p_z=p^R_z\end{matrix}}.
\end{align}

Then the nonperturbatively renormalized quasi-PDF can be matched to the PDF in the $\overline{\text{MS}}$ scheme:
\begin{align} \label{eq:momfact}
\tilde{q}_{v, R}^{\pi}(x,\tilde n\cdot P,\tilde\mu)&= \int_{0}^1 {dy\over y}C\left({x\over y},\frac{\tilde\mu}{\mu},\frac{\mu}{y\tilde n\cdot P}\right)q_{v, R}^{\pi}(y,\mu)\non\\
&+\mathcal{O}\left({m_\pi^2\over (\tilde n\cdot P)^2},{\Lambda_\text{QCD}^2\over (\tilde n\cdot P)^2}\right),
\end{align}
where $\tilde\mu$ and $\mu$ denote the renormalization or cutoff scale for the quasi-PDF and the PDF, respectively. $m_\pi$ is the pion mass. 
The matching can be carried out perturbatively. 
At one loop, we define
\begin{equation}
C(\xi,\bar\eta,\eta)=\delta(1-\xi)(1-\frac{\alpha_s}{2\pi}\delta C^{(1)})+\frac{\alpha_s}{2\pi}C^{(1)}(\xi,\bar\eta,\eta)
\end{equation} 
with $\delta C^{(1)}=\int_{-\infty}^{\infty} d\xi C^{(1)}(\xi)$. The matching kernel is the same as in Ref.~\cite{Chen:2018xof}.
Ideally, the continuum limit should be taken before matching such that lattice artifact can be removed and rotational symmetry can be recovered. However, only a single lattice spacing is used in this work.

For power corrections, the meson mass correction associated with the choice of Dirac matrix $\gamma^t$ is identical to that of the helicity distribution worked out in Ref.~\cite{Chen:2016utp}. The $\mathcal{O}({\Lambda_\text{QCD}^2/(\tilde n\cdot P)^2})$ correction is numerically rather small in the present case. The renormalization and matching for the Wilson line renormalization scheme is shown in the Appendix.

\section{Lattice calculation setup}  \label{sec:lattice}
In addition to the setup described in the Introduction, the gauge links are one step hypercubic(HYP)-smeared~\cite{Hasenfratz:2001hp} with the clover parameters tuned to recover the lowest pion mass of the staggered quarks in the sea~\cite{Rajan:2017lxk,Bhattacharya:2015wna,Bhattacharya:2015esa,Bhattacharya:2013ehc}. On these configurations, we calculate the time-independent, nonlocal (in space, chosen to be in the $z$ direction) correlators of a pion with a finite-$P_z$ boost
\begin{align}
\label{eq:qlat}
\tilde{h}_\text{lat}(z,P_z,a) =  \frac{P_z}{P_0}
  \left\langle \pi(\vec{P}) \right|
    \bar{\psi}(z) \Gamma \left( \prod_n U_z(n\hat{z})\right) \psi(0)
  \left| \pi(\vec{P}) \right\rangle,
\end{align}
where $U_z$ is a discrete gauge link in the $z$ direction, $\vec{P}=\{0,0,P_z\}$ is the momentum of the pion, and $\Gamma=\gamma^t$. {For each momentum, we use the sequential approach to calculate the three-point function with \{8,16,16,32\} thousand measurements for the smallest source-sink separation to the largest one, respectively.}

It is worthwhile to point out that for the valence quark PDF considered in the present paper, the imaginary part vanishes. The reason is that the imaginary part yields $q_f^\pi(x) + q_{\bar{f}}^\pi(x)$. For a charged pion, the isovector combination
$q_u^\pi(x) + q_{\bar{u}}^\pi(x) - [q_d^\pi(x) + q_{\bar{d}}^\pi(x)]$ vanishes since $q_u^\pi(x) = q_{\bar{d}}^\pi(x)$ and $q_{\bar{u}}^\pi(x) =q_d^\pi(x)$ due to isospin symmetry.

To reach larger pion momentum, the optimal Gaussian smearing parameter was chosen by varying the parameters. Then $\widetilde{h}_{\text{latt}}(z,P_z,a)$ is computed by performing one- and two-state fits using the model~\cite{Bhattacharya:2013ehc}

\begin{align}
\label{eq:c3ptfitform}
C^\text{2pt}(P_z,t_\text{sep}) &=
   |{\cal A}_0|^2 e^{-E_0t_\text{sep}}+ |{\cal A}_1|^2 e^{-E_1t_\text{sep}}, \nonumber\\
C^\text{3pt}_{\Gamma}(P_z,t,t_\text{sep}) &=
   |{\cal A}_0|^2 \langle 0 | \mathcal{O}_\Gamma | 0 \rangle  e^{-E_0t_\text{sep}} \nonumber\\
   &+|{\cal A}_1|^2 \langle 1 | \mathcal{O}_\Gamma | 1 \rangle  e^{-E_1t_\text{sep}} \nonumber\\
   &+{\cal A}_1{\cal A}_0^* \langle 1 | \mathcal{O}_\Gamma | 0 \rangle  e^{-E_1 (t_\text{sep}-t)} e^{-E_0 t} \nonumber\\
   &+{\cal A}_0{\cal A}_1^* \langle 0 | \mathcal{O}_\Gamma | 1 \rangle  e^{-E_0 (t_\text{sep}-t)} e^{-E_1 t} ,
\end{align}
where $t_{sep}$ is the source sink separation,  $t$ is the operator insertion time,
$E_0$ ($E_1$) is the ground- (excited-) state nucleon energy and ${\cal A}_0$ (${\cal A}_1$) is the overlapping and kinematic factor for the ground- (excited-) state hadron.

\begin{figure}[htbp]
\includegraphics[width=.45\textwidth]{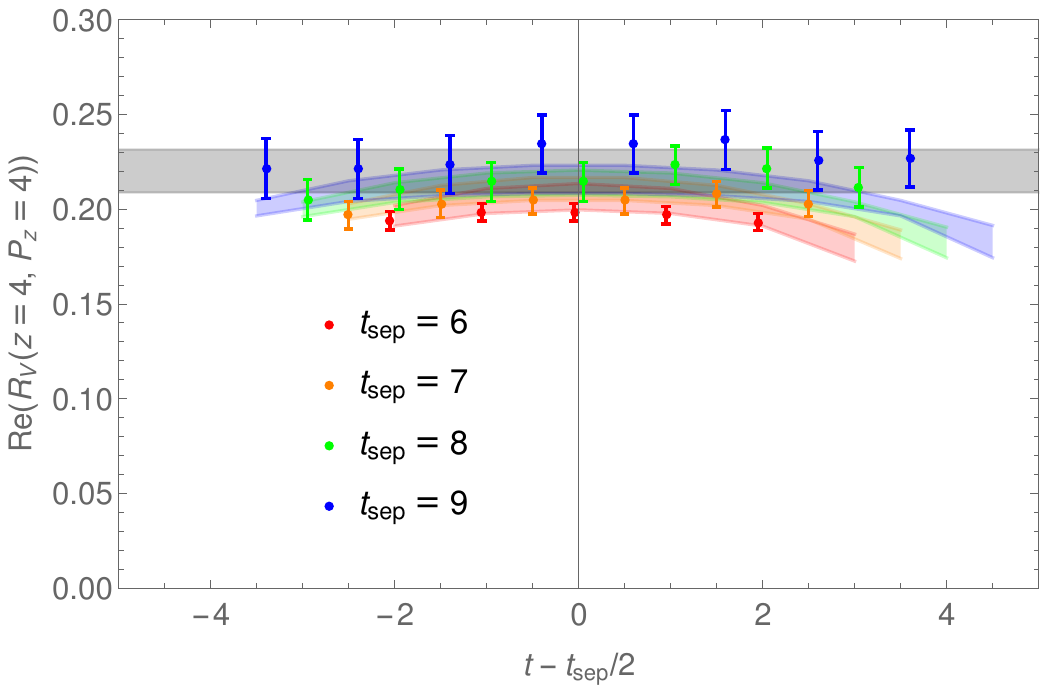}
\caption{Example plot of  the three- to two-point function ratios vs. the insertion time $t$ of the operator $ \mathcal{O}_\Gamma$. The real parts of the pion matrix elements are shown with pion momentum $P_z=4\frac{2\pi}{L}$ and length of the Wilson line $z=4$ in units of $a$(=0.12 fm). The curved bands are simultaneous fits to source sink separations $t_{\text{sep}}/a=6,7,8,9$ using Eq.~(\ref{eq:c3ptfitform}) but without the $\langle 1 | \mathcal{O}_\Gamma | 1 \rangle$ term. The straight horizontal bands are the extracted $\langle 0 | \mathcal{O}_\Gamma | 0 \rangle$.} 
\label{bareME-ratio}
\end{figure}

An example plot for $z/a=4$ with pion momentum $P_z =4\times 2\pi/L=1.74$~GeV are shown in Fig.~\ref{bareME-ratio}. The three- to two-point function ratio vs. the operator insertion time $t$ are shown. The straight horizontal band is the extracted $\langle 0 | \mathcal{O}_\Gamma | 0 \rangle$ from the simultaneous fits to source sink separation $t_{sep}/a=6,7,8,9$ using Eq.~(\ref{eq:c3ptfitform}) but with the term $\langle 1 | \mathcal{O}_\Gamma | 1 \rangle$ omitted. 

In Fig.~\ref{bareME-fitcomp}, we show the comparison between the one- and two-state fits. The one-state fit is performed with four different $t_\text{sep}$ ($t_\text{sep}/a=6,7,8,9$) while the two-state fit is performed with the same method as in Fig. 1. The excited state contamination is expected to be smaller when $t_\text{sep}$ is larger. This might be the cause of the shift of the central values as $t_\text{sep}$ increases although the shift is still within errors. The bands are two state fits as shown in Fig. (\ref{bareME-ratio}). The one and two state fits are consistent within errors.


In Fig.~\ref{fig:ME-tsep}, we plot the real {and imaginary} part of the $P_z=4\times 2\pi/L$ pion matrix elements renormalized with the RI/MOM scheme as shown in Eqs.(\ref{hR},\ref{hRx}) with $\mu_R=3.7$ GeV, $p_z^R=6\times 2\pi/L$, {where the imaginary part arises from the RI/MOM renormalization factor which is complex at nonzero $p_z^R$ and shall be viewed as a scheme dependence.}
The ``two-two'' (same fitting method as used in Fig. (\ref{bareME-ratio})) and ``two-twoRR'' (all the terms in Eq.~(\ref{eq:c3ptfitform}), including the $\langle 1 | \mathcal{O}_\Gamma | 1 \rangle$ term, are included in the fit) analyses use all four source-sink separations while the ``two-two2sep'' uses only the largest two source-sink separations. One can see that different two-state analyses are also consistent with each other.

The consistency of the one- and two-state fits with multiple $t_{sep}$ and $t$ suggests that the residual error from excited state contamination is within our errors. In the following, we will use matrix elements from the ``two-twoRR'' analysis in our PDF analysis.
We use multiple values of pion momenta, $P_z=\{0,0, n \frac{2\pi}{L}\}$, with 
$n \in \{2,3,4\}$, which correspond to 0.86, 1.32 and 1.74~GeV, respectively.

\section{Numerical results and discussions}  
Now we present our numerical results for the valence quark distribution in the pion and discuss their physical implications. We first Fourier transform the renormalized lattice data to momentum space
\begin{equation}\label{FThR}
\tilde q_R(x)=\int \frac{dz}{4\pi}e^{ix z P_z}\tilde h_R(z),
\end{equation}
form the valence distribution $\tilde q_R(x)+\tilde q_R(-x)$, and then apply one-loop matching and meson mass corrections. The meson mass corrections are numerically rather small. To illustrate the impact of one-loop matching, we show the results before and after applying the matching at the largest momentum $P_z=4\times 2\pi/L$ in Fig.~\ref{fig:xqvpi}. As can be seen from the plot, one-loop matching results in a sizable contribution and shifts of the original quasi-PDF towards the physical region $[0, 1]$ in both schemes.

In our earlier work~\cite{Lin:2017ani} on the nucleon PDF, we also proposed a ``derivative'' method to improve the truncation error in the Fourier transform in Eq.~(\ref{FThR}). We take the 
derivative of the renormalized nucleon matrix elements $\partial_z\tilde{h}_R(z)$, whose Fourier transform differs from the original matrix element in a known way:
\begin{equation}
\label{eq:derivative}
\tilde{q}_R(x) = \int \frac{dz}{4\pi} \frac{ie^{i x P_z z}}{ x P_z} \partial_z\tilde{h}_R(z),
\end{equation}
where the surface terms vanish provided that $\tilde{h}_R(z)$ goes to zero as $|z|\to\infty$. 
In practice, the integral in Eq.~(\ref{eq:derivative}) has to be truncated at $|z|=|z_{\rm max}|$ due to finite lattice volume, implying that the contribution from long-range correlation inside the integral is cut off. Such contribution is expected to be small for large $P_z$ (to be more precise, for large $x P_z$) due to the oscillating phase.
We also applied this method to the present lattice data. {In Fig.~\ref{fig:xqwandwoderiv}, we show a comparison between the results with and without using the ``derivative" method. As can be seen from the figure, the two results are consistent with each other within errors, while the one with ``derivative" method exhibits less oscillating behavior.}

\begin{widetext}

\begin{figure}[htbp]
\includegraphics[width=.9\textwidth]{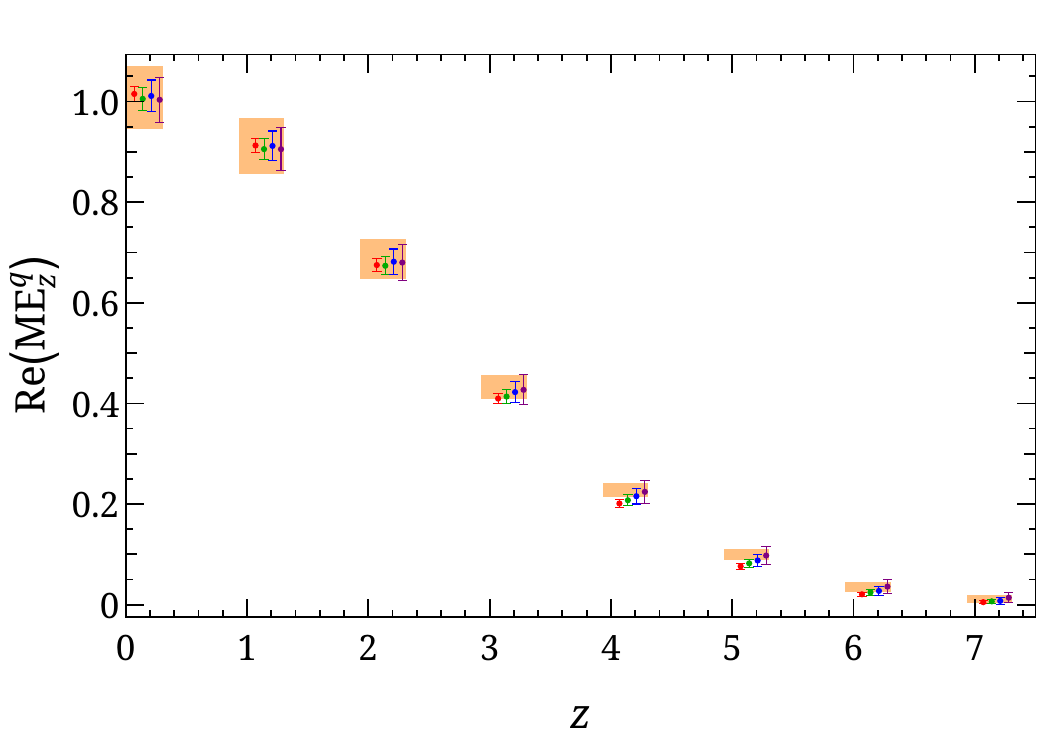}
\caption{Comparison between one state and two state fits.
The data points from left to right indicate the single-state fits (i.e. keeping only the $\langle 0 | \mathcal{O}_\Gamma | 0 \rangle$ terms in two- and three- point functions in Eq.~(\ref{eq:c3ptfitform}) for $t_\text{sep}/a=6,7,8,9$. The excited state contamination is expected to be smaller when $t_\text{sep}$ is larger. This might be the cause of the shift of the central values as $t_\text{sep}$ increases although the shift is still within errors. The bands are two state fits as shown in 
Fig. (\ref{bareME-ratio}). The one and two state fits are consistent within errors. 
} 
\label{bareME-fitcomp}
\end{figure}

\begin{figure}[htbp]
\includegraphics[width=.45\textwidth]{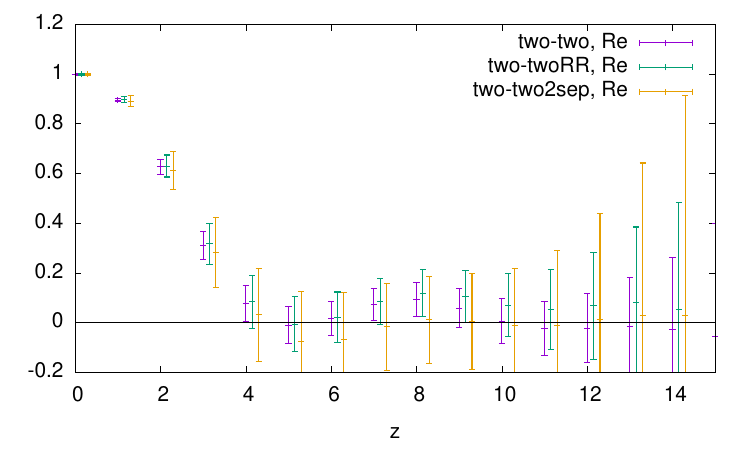}
\includegraphics[width=.45\textwidth]{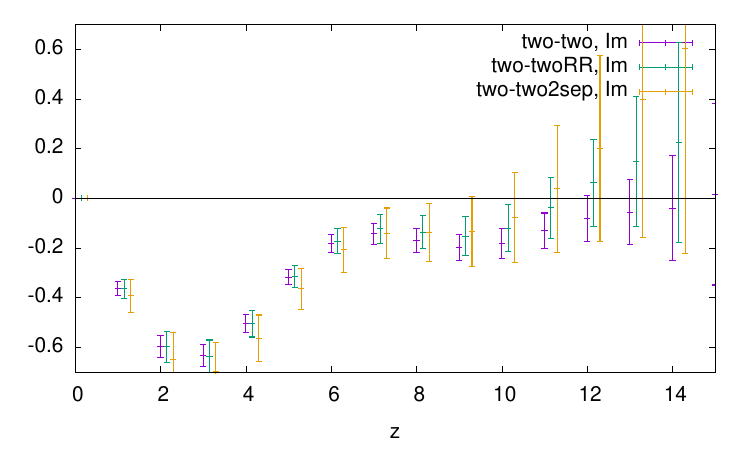}
\caption{The real {and imaginary} part of the $P_z=4\times 2\pi/L$ pion matrix elements renormalized with the RI/MOM scheme as shown in Eqs.(\ref{hR},\ref{hRx}) with $\mu_R=3.7$ GeV, $p_z^R=6\times 2\pi/L$.
The ``two-two'' (same fitting method as used in Fig. (\ref{bareME-ratio})) and ``two-twoRR'' (all the terms in Eq.~(\ref{eq:c3ptfitform}), including the $\langle 1 | \mathcal{O}_\Gamma | 1 \rangle$ term, are included in the fit) analyses use all four source-sink separations while the ``two-two2sep'' uses only the largest two source-sink separations.  }
\label{fig:ME-tsep}
\end{figure}

\end{widetext}

\begin{figure}[tbp]
\includegraphics[width=.5\textwidth]{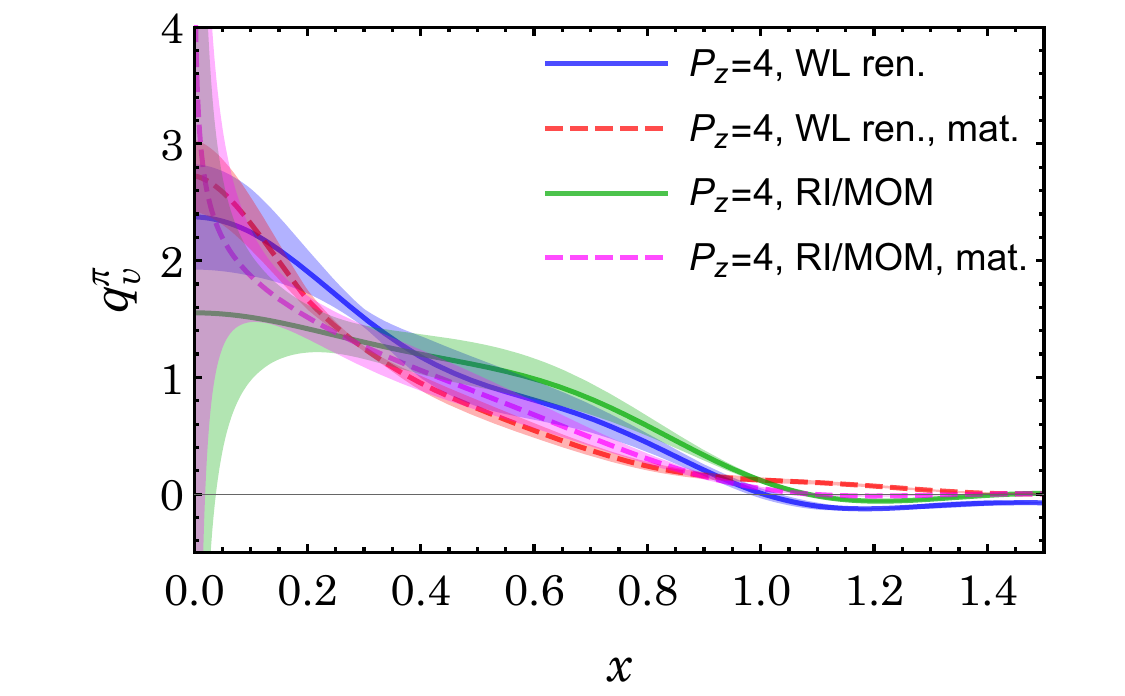}
\caption{The pion valence quark PDF result from the Fourier transform in \Eq{FThR} (blue for Wilson line renormalization and green for RI/MOM scheme), after one-loop matching (red dashed for Wilson line renormalization and purple dashed for RI/MOM scheme), for the momentum $P_z=4\times2\pi/L$.} \label{fig:xqvpi}
\end{figure}

\begin{figure}[tbp]
\includegraphics[width=.5\textwidth]{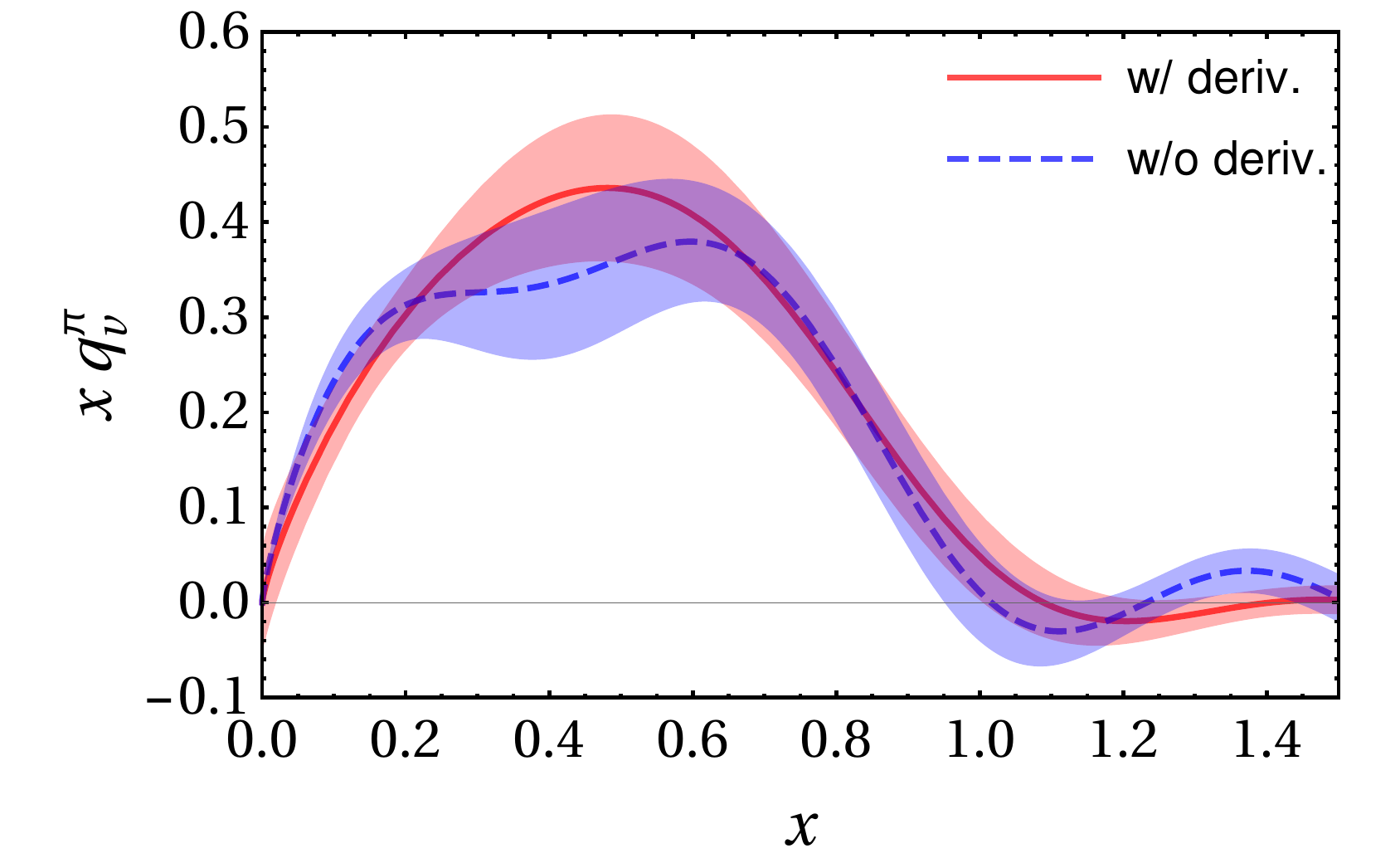}
\caption{{Comparison between the results with (red) and without (blue) using the ``derivative" method, for the momentum $P_z=4\times2\pi/L$, and $\mu_R=3.7$ GeV, $p_z^R=6\times2\pi/L$. Both results include statistical errors as well as systematic error due to $p_z^R$ dependence by varying it between $4$ and $8$.}} \label{fig:xqwandwoderiv}
\end{figure}

\begin{figure}[tbp]
\includegraphics[width=.5\textwidth]{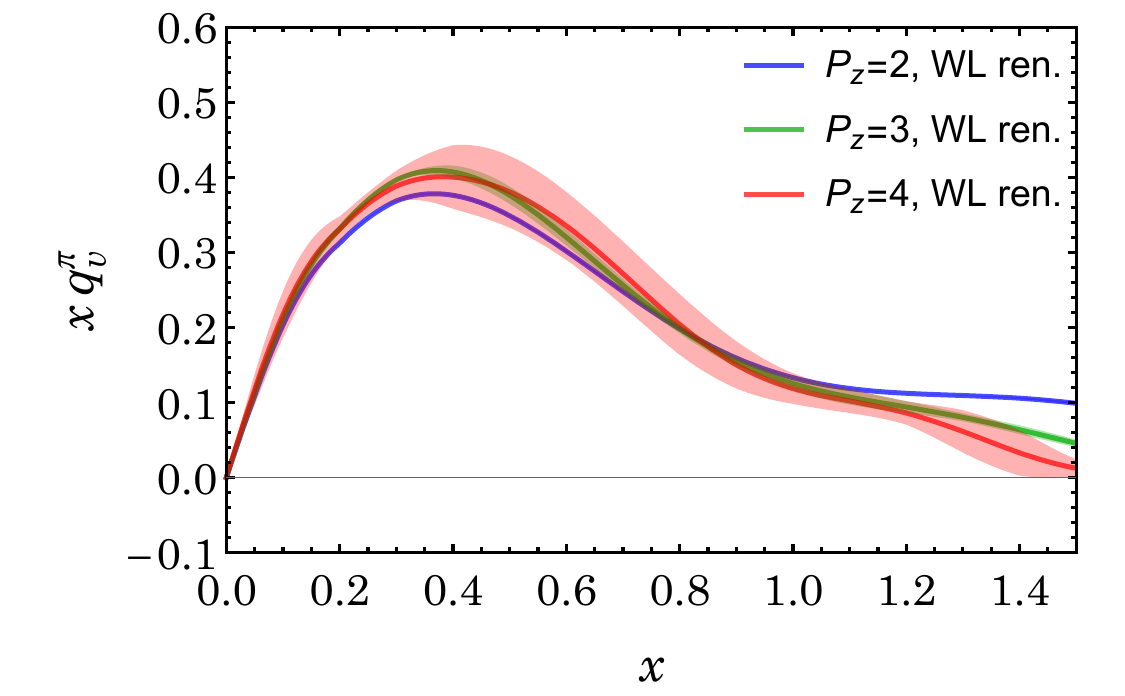}
\includegraphics[width=.5\textwidth]{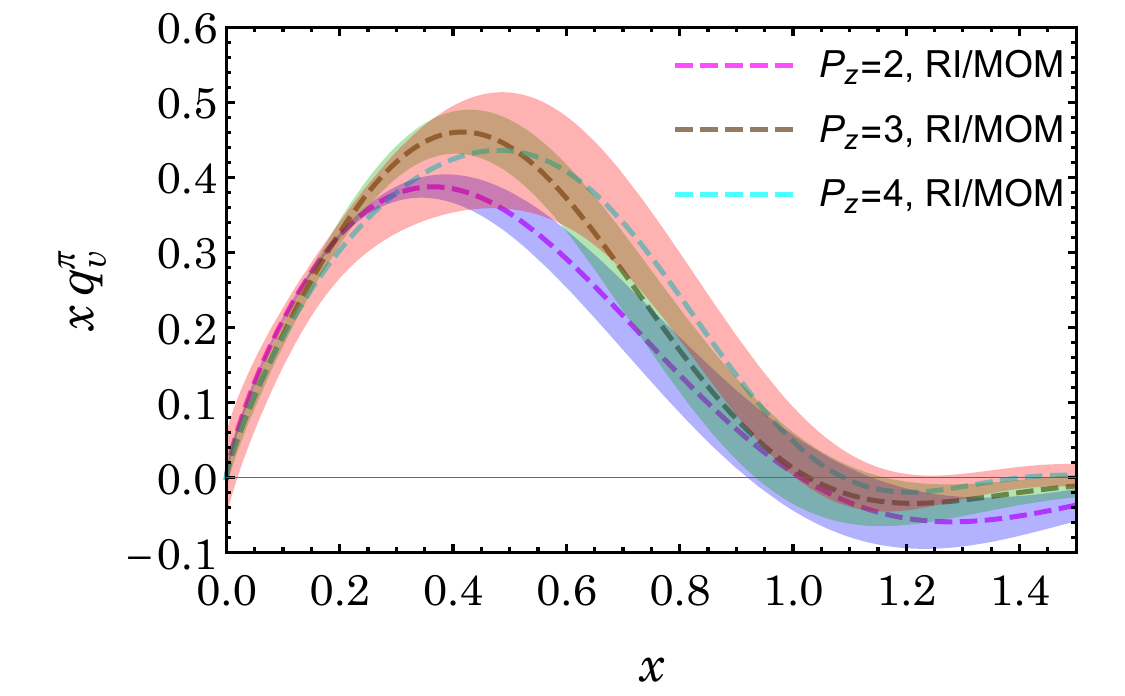}
\caption{The pion momentum dependence of the results in the Wilson line renormalization scheme (WL ren.) and RI/MOM scheme. We have chosen $\mu_R=3.7$ GeV and $p_z^R=6\times2\pi/L$ in the RI/MOM result. The Wl ren. result includes statistical errors, whereas the RI/MOM result also includes systematic error due to $p_z^R$ dependence by varying it between $4$ and $8$.}  \label{fig:qvpzcomp}
\end{figure}

\begin{figure}[tbp]
\includegraphics[width=.5\textwidth]{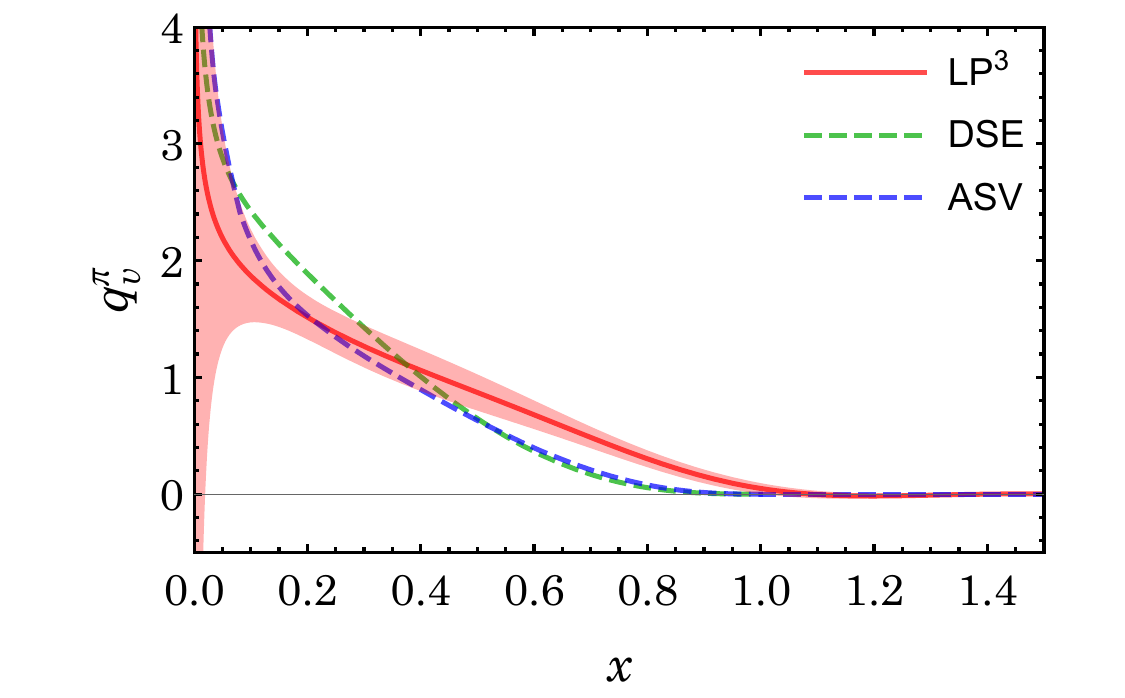}
\includegraphics[width=.5\textwidth]{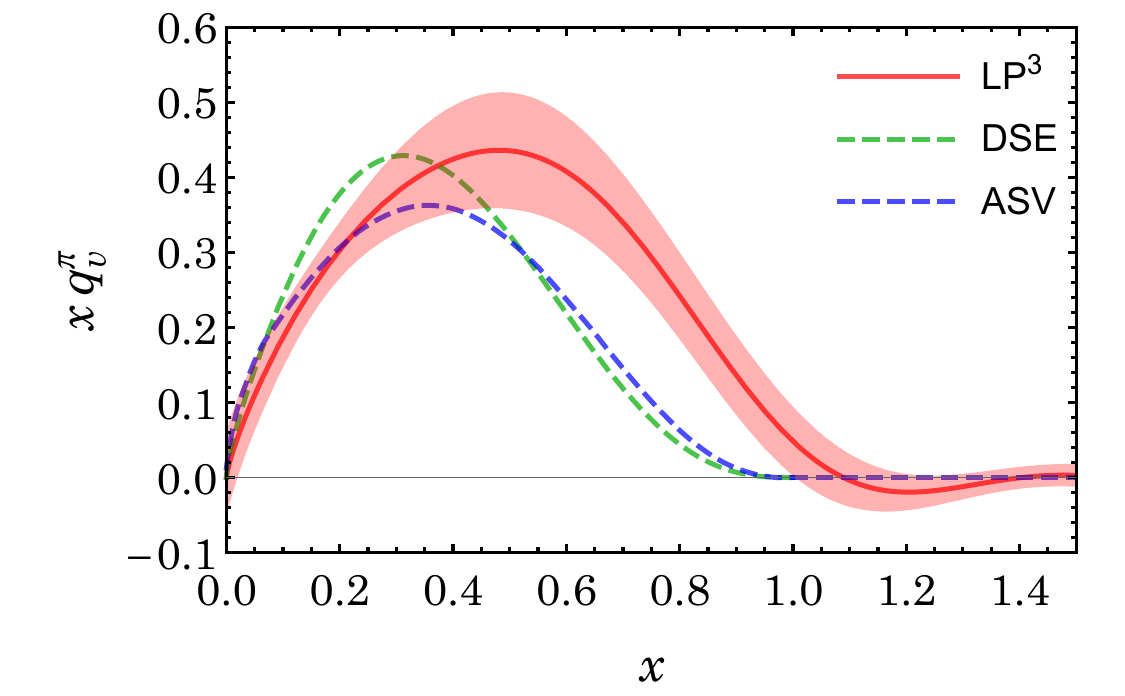}
\caption{Our pion valence-quark PDF result at the scale $\mu=4$ GeV from RI/MOM scheme calculation (LP$^3$) , contrasted with analysis from Dyson-Schwinger equation~\cite{Chen:2016sno} (DSE) at the scale $5.2$ GeV and from a fit to Drell-Yan data in Ref.~\cite{Aicher:2010cb} (ASV) at $4$ GeV.} \label{fig:qvDphenocomp}
\end{figure}

In Fig.~\ref{fig:qvpzcomp}, we show the results of the valence quark distribution for different pion momenta and renormalization schemes. For the RI/MOM result, we have chosen $\mu_R=3.7$ GeV, $p_z^R=6\times 2\pi/L$, and included statistical as well as the systematic error of setting the unphysical scale $p_z^R$ by varying it between $4\times 2\pi/L$ and $8\times 2\pi/L$. For the Wilson line renormalization, only statistical errors are included since there is no extra unphysical scale in this scheme like $p_z^R$ in RI/MOM. As can be seen from the figure, increasing $P_z$ tends to shift the distribution towards $x=0$ and also lifts the peak at $x=0$, but its impact is mild. 
Another important feature is that the RI/MOM result is consistent with $0$ outside the physical region $[0,1]$ within errors, whereas the Wilson line renormalization one is not. This mainly reflects the importance of the higher-order matching kernel, since the one-loop matching in the two different schemes differ only by finite terms. We plan to derive higher-order matching, expecting that it will reduce the difference between the results in two different schemes. 

It is worthwhile to stress that matching is a necessary step in converting the quasi-PDF to PDF. It yields sizeable contributions and changes, in particular for the distribution in the unphysical region. In Ref.~\cite{Xu:2018eii}, the authors studied the pion valence quasi-distribution using the Bethe-Salpeter wave function of the pion, and observed that for $P_z\gtrsim 2$ GeV, by further increasing the pion momentum the quasi-PDF shrinks to the physical region very slowly. Actually we have observed a similar trend in our data. However, the matching plays an important role in reducing the contribution in the unphysical region, as can be seen from Fig.~\ref{fig:xqvpi} above, but hasn't been taken into account in Ref.~\cite{Xu:2018eii}. 

In Fig.~\ref{fig:qvDphenocomp}, we compare our final result in RI/MOM scheme ($\textrm{LP}^3$) with computations from Dyson-Schwinger equation~\cite{Chen:2016sno} (DSE) and from a phenomenological fit to Drell-Yan data~\cite{Aicher:2010cb} (ASV), {where our error band includes statistical and
systematic error of setting the unphysical scale $p_z^R$, as was done in Fig. 3.}
We have set our renormalization scale to be $\mu=4$~GeV, in accordance with the experimental fit~\cite{Aicher:2010cb}, whereas the DSE result is at $5.2$~GeV. 
Outside the physical region, our result is consistent with $0$. Within the physical region, our result decreases more slowly than the DSE and ASV results at large $x$, and has a lower peak around $x=0$, as can be seen from the upper plot. This is expected to improve once we have lattice data at smaller pion masses. When plotted as $x q_v^{\pi}(x)$, as was usually done in the literature, the discrepancy at small $x$ gets suppressed, while it gets enhanced at large $x$.

We point out several potential sources of uncertainty or artifact in the above analysis, which we aim to improve in the future. First, the contribution at large $x$ depends on the pion momentum as well as on the unphysical pion mass used in this calculation. If we have a larger pion momentum and a pion mass closer to its physical value, the contribution at large $x$ will be further reduced, and accordingly, the small $x$ contribution will be enhanced. Second, the matching implemented here is at one-loop order. It has a sizable effect and shifts the result towards the physical region. It is therefore important to investigate the impact of higher-order matching, in order to reduce uncertainties due to perturbative matching. This is also reflected by the difference between the results in two different schemes. Third, the present calculation is carried out at one lattice spacing, we'll need data at more lattice spacings to have a continuum extrapolation. 
Last but not least, we also need simulations with larger volumes to control the finite volume effect.

\section*{Acknowledgments}
We thank the MILC Collaboration for sharing the lattices used to perform this study. {We also thank Yu-Sheng Liu and Yi-Bo Yang for providing part of the data/plots and useful discussions.} The LQCD calculations were performed using the Chroma software suite~\cite{Edwards:2004sx}. 
This research used resources of the National Energy Research Scientific Computing Center, a DOE Office of Science User Facility supported by the Office of Science of the U.S. Department of Energy under Contract No. DE-AC02-05CH11231 through ALCC and ERCAP; facilities of the USQCD Collaboration, which are funded by the Office of Science of the U.S. Department of Energy, and supported in part by Michigan State University through computational resources provided by the Institute for Cyber-Enabled Research. JWC is partly supported by the Ministry of Science and Technology, Taiwan, under Grant No. 105-2112-M-002-017-MY3 and the Kenda Foundation.  LCJ is supported by the Department \\ of Energy, Laboratory Directed Research and Development (LDRD) funding of BNL, under contract DE-EC0012704. HL is supported by the US National Science Foundation under grant PHY 1653405 ``CAREER: Constraining Parton Distribution Functions for New-Physics Searches''. AS and JHZ are supported by the SFB/TRR-55 grant ``Hadron Physics from Lattice QCD''. JHZ is also supported by a grant from National Science Foundation of China (No.~11405104). YZ is supported by the U.S. Department of Energy, Office of Science, Office of Nuclear Physics, from DE-SC0011090 and within the framework of the TMD Topical Collaboration.

\vspace*{2em}

\section*{Appendix A: Wilson Line Renormalization Scheme}

In this scheme, the nonperturbative renormalization reads
\begin{equation}\label{hrenorm}
\tilde h_R(\lambda\tilde n)=Z_1 Z_2 e^{\delta m\lambda}\tilde h(\lambda\tilde n),
\end{equation}
The counterterm $\delta m$ has been extracted nonperturbatively on the lattice with $\delta m=253(3)$ MeV at $a=0.12$~fm
~\cite{Zhang:2017bzy,Chen:2017gck}. The renormalization factors $Z_1$ and $Z_2$ come from the endpoint renormalization of $\tilde h(\lambda\tilde n)$ and can be fixed by an overall normalization. 

The matching kernel $C^{(1)}$ reads 
\begin{widetext}
\begin{align}\label{unpolvertexnoexp}
C^{(1)}(x)/C_F&=\left\{ \begin{array} {ll} \frac{1+x^2}{1-x}\ln \frac{x(\Lambda(x)-x P_z)}{(x-1)(\Lambda(1-x)+P_z(1-x))}+1+\frac{\Lambda(1-x)-\Lambda(x)}{P_z}+\frac{x\Lambda(1-x)+(1-x)\Lambda(x)}{(1-x)^2 P_z}-\frac{\Lambda}{(1-x)^2P_z}\ , & x>1\ , \\
\ \\
\frac{1+x^2}{1-x}\ln\frac{(P_z)^2}{\mu^2}+\frac{1+x^2}{1-x}\ln \frac{4x(1-x)(\Lambda(x)-x P_z)}{\Lambda(1-x)+(1-x)P_z}-\frac{2}{1-x}+1+2x+\frac{\Lambda(1-x)-\Lambda(x)}{P_z}\ & \\
+\frac{x\Lambda(1-x)+(1-x)\Lambda(x)}{(1-x)^2 P_z}-\frac{\Lambda}{(1-x)^2P_z}\ , & 0<x<1\ , \\
\ \\
\frac{1+x^2}{1-x}\ln \frac{(x-1)(\Lambda(x)-x P_z)}{x(\Lambda(1-x)+(1-x)P_z)}-1+\frac{\Lambda(1-x)-\Lambda(x)}{P_z}+\frac{x\Lambda(1-x)+(1-x)\Lambda(x)}{(1-x)^2P_z}-\frac{\Lambda}{(1-x)^2P_z}\ , & x<0 \ ,\end{array} \right.
\end{align}
%
where $\Lambda(x) = \sqrt{\Lambda^2+x^2 P_z^2}$ with $\Lambda$ being a transverse momentum cutoff. 
\end{widetext}

\bibliography{bibliography}
\end{document}